\documentclass[letterpaper, 10 pt, conference]{ieeeconf}  
\IEEEoverridecommandlockouts
\usepackage[noadjust]{cite}
\usepackage{amsmath,amssymb,amsfonts}
\usepackage{algorithmic}
\usepackage{subfigure}
\usepackage{graphicx}
\usepackage{textcomp}
\usepackage{xcolor}
\usepackage{soul}
\usepackage{graphics} 
\usepackage{epsfig} 
\usepackage{mathptmx} 
\usepackage{times} 
\usepackage{amsmath} 
\usepackage{amssymb}  
\usepackage{multirow}
\usepackage{graphicx, epstopdf}
\usepackage{xspace,epsfig,url}
\usepackage{psfrag}
\usepackage{verbatim}
\usepackage[all]{xy}
\usepackage{color}
\usepackage{comment}
\usepackage{cases}
\usepackage{hyperref}

\def\BibTeX{{\rm B\kern-.05em{\sc i\kern-.025em b}\kern-.08em
    T\kern-.1667em\lower.7ex\hbox{E}\kern-.125emX}}
  
\begin{document}


\title{\LARGE \bf
Effects of Haptic Feedback on the Wrist\\ during Virtual Manipulation
}

\author{Mine Sarac$^{1}$, Allison M. Okamura$^{1}$, and Massimiliano Di Luca$^{2}$
\thanks{$^{1}$Department of Mechanical Engineering, Stanford University,  Stanford, CA.
        {\tt\small msarac@stanford.edu, aokamura@stanford.edu}}%
\thanks{$^{2}$Previously at Facebook Reality Labs and now at University of Birmingham, UK
        {\tt\small m.diluca@bham.ac.uk}}%
}


\maketitle
\thispagestyle{empty}
\pagestyle{empty}

\begin{abstract}
We propose a haptic system for virtual manipulation to provide feedback on the user’s forearm instead of the fingertips. In addition to visual rendering of the manipulation with virtual fingertips, we employ a device to deliver normal or shear skin-stretch at multiple points near the wrist. To understand how design parameters influence the experience, we investigated the effect of the number and location of sensory feedback on stiffness perception. Participants compared stiffness values of virtual objects, while the haptic bracelet provided interaction feedback on the dorsal, ventral, or both sides of the wrist. Stiffness discrimination judgments and duration, as well as qualitative results collected verbally, indicate no significant difference in stiffness perception with stimulation at different and multiple locations.

\end{abstract}


\section{Introduction}

Kinesthetic haptic devices often aim to recreate mechanical properties of objects such as mass, stiffness, and temperature, during virtual manipulation tasks.
Fingertip devices~\cite{Suchoski2018, Leonardis2017} can provide realistic sensations, but their design is complex, and for some applications, such as augmented reality tasks, it is desirable to leave the fingertips free to interact with physical objects. Haptic bracelets (or arm bands) can overcome these issues by relocating the interaction forces to the wrist and leaving the fingertips free \cite{Pacchierotti2017}. 

Some haptic bracelets provide squeeze the user's wrist 
in a distributed manner~\cite{Pezent2019, Young2019, RaitorICRA2017}, and help users perform better in virtual tasks compared to not having haptic feedback. However, they do not focus on the effect of the number or the location of contact points. Moriyama~\textit{et al.}~\cite{Moriyama2018} compared the ``strange'' feeling of the feedback on the ventral and dorsal side of the wrist, and found no significant difference.  


In this work, we focus on comparing the feedback on different locations or different numbers of contact points in terms of task performance. We design a palpation study, where users push and press virtual objects, 
and receive haptic feedback on the dorsal, ventral, or both sides of the wrist. 


\section{Experimental Setup}

\begin{figure}[t]
  \centering
  \resizebox{2.8in}{!}{\includegraphics{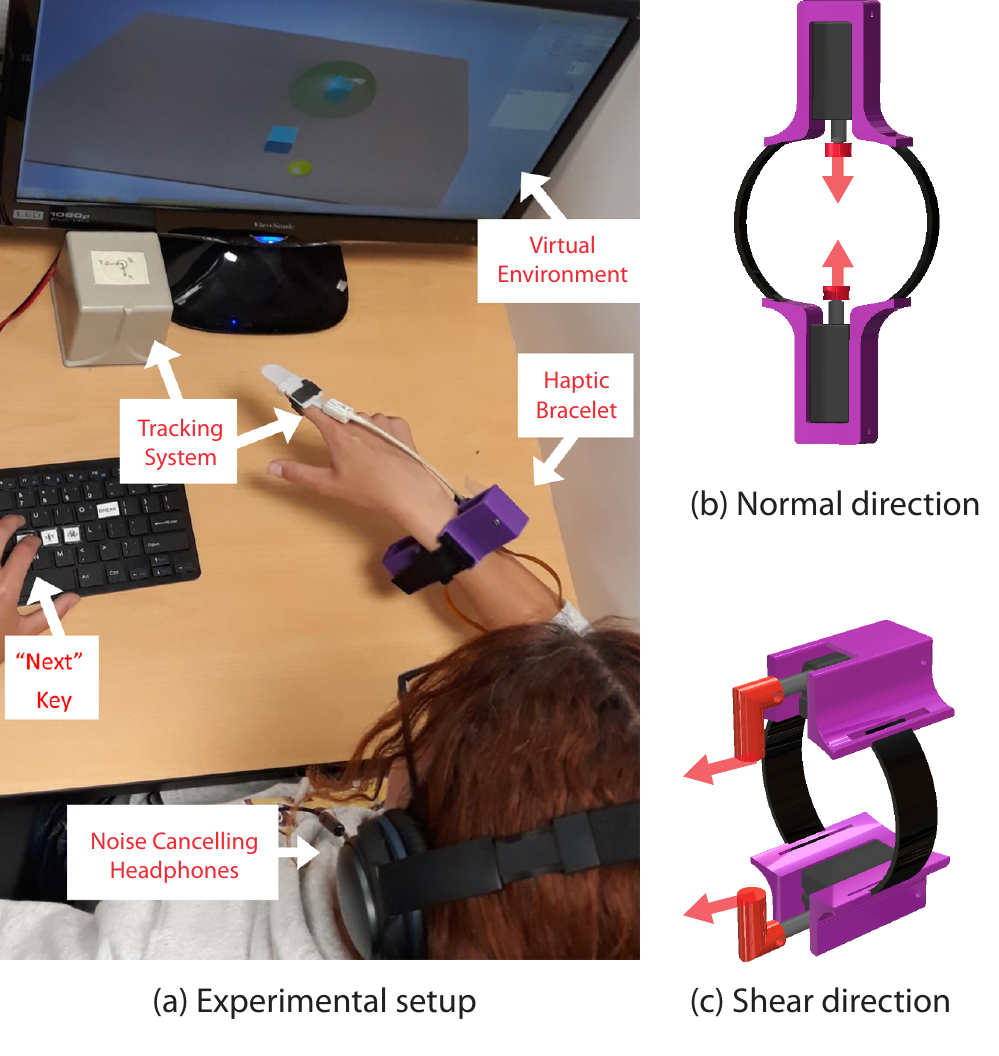}}
  \vspace*{-.75\baselineskip}
  \caption{(a) Experiment setup: A user wears the haptic bracelet, a fingertip sensor for tracking, and noise cancelling headphones. As users interact with virtual objects on the monitor, they receive haptic cues accordingly in the direction of (b) normal and (c) stretch/shear on the dorsal, ventral, or both sides of the wrist. Slip is prevented using a double-sided tape for shear.}
  \label{fig:setup}
  \vspace*{-1.25\baselineskip}
\end{figure}

Fig.~\ref{fig:setup}(a) shows the experiment setup with a haptic bracelet, a virtual environment, and a tracking system. We designed two bracelets with identical linear actuators (Actuonix PQ12-P) and different groundings to apply feedback in the normal (Fig.~\ref{fig:setup}(b)) and shear (Fig.~\ref{fig:setup}(c)) directions, because haptic bracelets in the literature apply normal or shear forces \cite{Aggravi2018}. 
As the virtual interaction occurs, we moved the actuator on the dorsal, ventral, or both sides of the wrist.

The virtual environment is created with the CHAI3D framework~\cite{Conti2005} and displayed on a monitor at 144~Hz. 
The user's finger movements are tracked with a trakSTAR system and a sensor grounded on the user's finger. 

During the experiment, users were asked to interact with two visually identical box objects with different simulated stiffness values (0.3 vs. 0.1, 0.2, 0.4, or 0.5~N/mm) and to choose the stiffer one. Users were shown whether their answer was correct in training mode, but not in testing mode.

We recruited 12 volunteers to participate in the study. The order of the conditions were randomized for each user.  
The overall experiment was composed of two parts, one for each feedback direction. Each part had 1 training block with 24 trials, and 3 testing blocks with 16 trials each. Each testing block rendered forces from one location of force, while the training block used all of the locations in a predefined order. Once the first part was completed, the user donned the other bracelet and repeated the procedure. Thus, each user experienced both feedback directions. Between each block, the user rested as needed. 

\section{Results and Discussion}

\begin{figure}[b!]
 \vspace*{-1.\baselineskip}
\centering
{\includegraphics[width=0.49\textwidth]{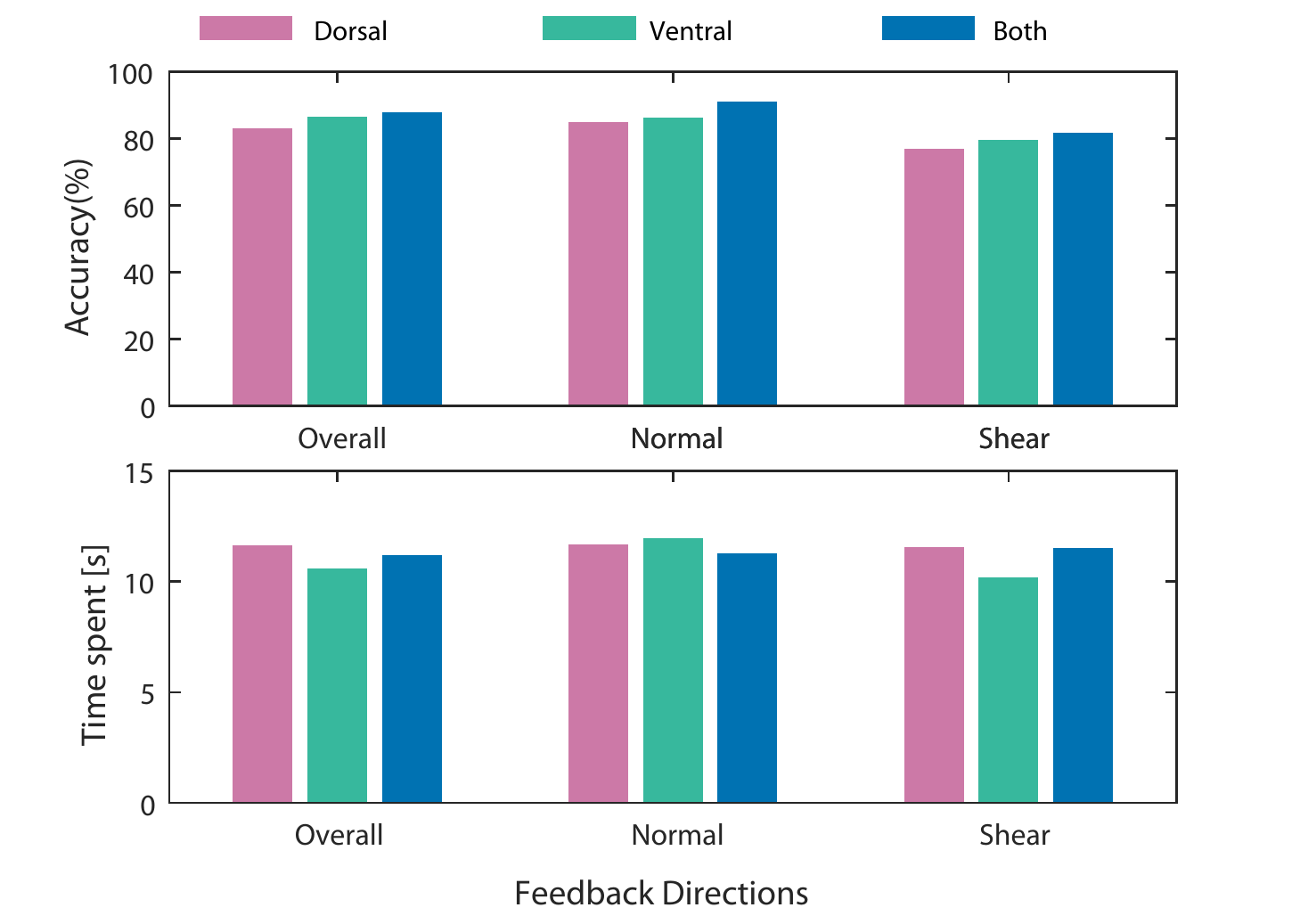}}
 \vspace*{-1.25\baselineskip}
\caption{Average accuracy of users' responses and time spent based on feedback location for all trials, for trials with feedback in normal or shear directions. Even though a similar trend is observed with different feedback directions, no statistical significance was found in accuracy.}
	\label{fig:LocationComparison}
\end{figure}

Most users reported that they were comfortable receiving the relocated feedback from the beginning, while others became comfortable over time. Fig.~\ref{fig:LocationComparison} compares the average accuracy of users' responses for the dorsal, ventral, and both contact locations. The data is shown for all trials, for trials in the normal direction only, and in the shear direction only. 

We performed a repeated measures ANOVA to analyze the data based on location and number of contact points. We found that the performance of users was significantly different 
in the normal direction only [F(2,22) = 4.9699, p = 0.01655], but not for the shear direction [F(2,22) = 0.7633, p = 0.7633], nor overall [F(2,22) = 2.7547, p = 0.0855]. On the other hand, users' average time to complete the trials was significantly different 
in the shear direction only [F(2,22) = 4.8062, p = 0.01854], but not for the normal direction [F(2,22) = 0.2167, p = 0.8068], nor overall [F(2,22) = 0.8409, p = 0.4447].  
 

We asked users to choose the feedback location they preferred. 1 user reported ventral, 2 reported dorsal, 5 reported both sides, and 3 users reported no difference. Then, we asked them to choose the most noticeable feedback location. 3 users reported ventral, 5 reported dorsal, and 4 reported both sides. The exact phrasing used in such an evaluation, such as the ``strange feeling’’ question from~\cite{Moriyama2018}, might affect user response. Furthermore, we found no correlation between the location in which they performed the best and the location in which they \emph{thought} they performed the best. 

\section{Conclusion and Future Work}

In this work, we analyzed the effects of location and number of contact points for haptic bracelets in terms of virtual objects' perception. We compared feedback on the dorsal, ventral, and both sides of the wrist, while the feedback was given in the normal or shear direction. Even though users tend to perform the best when the haptic feedback is given on both sides and the worst with feedback is given on dorsal side only, they are not significantly different. Given these results, the best location might be subjective and not very important in terms of perception, as long as the feedback is rendered on user's skin effectively.

Fig.~\ref{fig:LocationComparison} also shows that users perform better when the feedback is given in the normal direction than shear. However, this result might be caused by the fact that the actuators are moved an equal distance in each direction, causing the rendered forces around the wrist to be lower in the shear direction than the normal direction due to differing skin stiffness. Although Diller~\textit{et al.}~\cite{Diller2001} measured skin stiffness around the wrist, we experienced that using different equipment in terms of contact area, wearability, or the given displacement of the actuators might change these measurements. The difference in skin stiffness makes it harder to compare the normal and shear directions. In particular, Moriyama~\textit{et al.} reported the superiority of feedback in the normal direction in terms of ``strange'' feeling, and Biggs et al.~\cite{Biggs2002} reported that subjects are more sensitive in the shear direction compared to the normal. In future work, we will run a method of adjustments study to find the actuator displacements in different directions with similar intensity on the user's wrist and compare the effect of these directions in terms of perception of mechanical properties.

In addition, users mostly interact with virtual objects using multiple fingers instead of a single finger. For this study, we simplified the task to the index finger to create a consistent, repeatable task. In the future, we will extend our analyses to two-finger grasping and identification of other mechanical properties using the lessons learned from this study. 



\bibliographystyle{IEEEtran}
\bibliography{HapticBracelet_HapticsSymp}

\begin{thebibliography}{10}
\providecommand{\url}[1]{#1}
\csname url@rmstyle\endcsname
\providecommand{\newblock}{\relax}
\providecommand{\bibinfo}[2]{#2}
\providecommand\BIBentrySTDinterwordspacing{\spaceskip=0pt\relax}
\providecommand\BIBentryALTinterwordstretchfactor{4}
\providecommand\BIBentryALTinterwordspacing{\spaceskip=\fontdimen2\font plus
\BIBentryALTinterwordstretchfactor\fontdimen3\font minus
  \fontdimen4\font\relax}
\providecommand\BIBforeignlanguage[2]{{%
\expandafter\ifx\csname l@#1\endcsname\relax
\typeout{** WARNING: IEEEtran.bst: No hyphenation pattern has been}%
\typeout{** loaded for the language `#1'. Using the pattern for}%
\typeout{** the default language instead.}%
\else
\language=\csname l@#1\endcsname
\fi
#2}}

\bibitem{Suchoski2018}
J.~M. {Suchoski}, S.~{Martinez}, and A.~M. {Okamura}, ``Scaling inertial forces
  to alter weight perception in virtual reality,'' in \emph{2018 IEEE
  International Conference on Robotics and Automation (ICRA)}, 2018, pp.
  484--489.

\bibitem{Leonardis2017}
D.~{Leonardis}, M.~{Solazzi}, I.~{Bortone}, and A.~{Frisoli}, ``A {3-RSR}
  haptic wearable device for rendering fingertip contact forces,'' \emph{IEEE
  Transactions on Haptics}, vol.~10, no.~3, pp. 305--316, 2017.

\bibitem{Pacchierotti2017}
C.~{Pacchierotti}, S.~{Sinclair}, M.~{Solazzi}, A.~{Frisoli}, V.~{Hayward}, and
  D.~{Prattichizzo}, ``Wearable haptic systems for the fingertip and the hand:
  {T}axonomy, review, and perspectives,'' \emph{IEEE Transactions on Haptics},
  vol.~10, no.~4, pp. 580--600, 2017.

\bibitem{Pezent2019}
E.~Pezent, A.~Israr, M.~Samad, S.~Robinson, P.~Agarwal, H.~Benko, and
  N.~Colonnese, ``Tasbi: {M}ultisensory squeeze and vibrotactile wrist haptics
  for augmented and virtual reality,'' in \emph{IEEE World Haptics Conference
  (WHC)}, 2019, pp. 1--6.

\bibitem{Young2019}
E.~M. Young, A.~H. Memar, P.~Agarwal, and N.~Colonnese, ``Bellowband: {A}
  pneumatic wristband for delivering local pressure and vibration,'' in
  \emph{IEEE World Haptics Conference (WHC)}, 2019, pp. 55--60.

\bibitem{RaitorICRA2017}
M.~Raitor, J.~M. Walker, A.~M. Okamura, and H.~Culbertson, ``{WRAP: W}earable,
  restricted-aperture pneumatics for haptic guidance,'' in \emph{IEEE
  International Conference on Robotics and Automation (ICRA)}, 2017, pp.
  427--432.

\bibitem{Moriyama2018}
T.~K. {Moriyama}, A.~{Nishi}, R.~{Sakuragi}, T.~{Nakamura}, and H.~{Kajimoto},
  ``Development of a wearable haptic device that presents haptics sensation of
  the finger pad to the forearm,'' in \emph{IEEE Haptics Symposium}, 2018, pp.
  180--185.

\bibitem{Aggravi2018}
M.~{Aggravi}, F.~{Pausé}, P.~R. {Giordano}, and C.~{Pacchierotti}, ``Design
  and evaluation of a wearable haptic device for skin stretch, pressure, and
  vibrotactile stimuli,'' \emph{IEEE Robotics and Automation Letters}, vol.~3,
  no.~3, pp. 2166--2173, 2018.

\bibitem{Conti2005}
F.~Conti, F.~Barbagli, D.~Morris, and C.~Sewell, ``{CHAI}: {A}n open-source
  library for the rapid development of haptic scenes,'' in \emph{IEEE World
  Haptics Conference (WHC)}, 2005, pp. 21--29.

\bibitem{Diller2001}
T.~T. Diller, ``Frequency response of human skin in vivo to mechanical
  stimulation,'' Master's thesis, Massachusetts Institute of Technology, 2001.

\bibitem{Biggs2002}
J.~{Biggs} and M.~A. {Srinivasan}, ``Tangential versus normal displacements of
  skin: {R}elative effectiveness for producing tactile sensations,'' in
  \emph{Proceedings of Symposium on Haptic Interfaces for Virtual Environment
  and Teleoperator Systems}, 2002, pp. 121--128.

\end{thebibliography}



\end{document}